# Radiation Pressure on a Diffractive Sailcraft

## GROVER A. SWARTZLANDER, JR.

*Center for Imaging Science, Rochester Institute of Technology, Rochester, NY 14623*
*Corresponding author: grover.swartzlander@gmail.com*

**Advanced diffractive films may afford advantages over passive reflective surfaces for a variety space missions that use solar or laser in-space propulsion. Three cases are compared: Sun-facing diffractive sails, Littrow diffraction configurations, and conventional reflective sails. A simple Earth-to-Mars orbit transfer at a constant attitude with respect to the sun-line finds no penalty for transparent diffractive sails. Advantages of the latter approach include actively controlled metasails and the reuse of photons.**

## 1. INTRODUCTION

Radiation pressure is a process whereby optical momentum is transferred to matter by means of reflection, diffraction, or absorption. This effect, first predicted by Maxwell [1,2], is strong enough to impart force and torque on satellites owing to the pressure from sunlight [3: Mariner 10, 1974], as well as on microscopic-scale bodies on Earth by use of a concentrated beam of light. In the latter case the object is generally trapped in two or three spatial dimensions owing to the use of a focused beam. Pure optical levitation may be achieved when a uniform beam of light is used, thereby affording free movement, as occurs with in-space solar radiation pressure. A rich body of scientific literature spanning more than nine decades [4,5] describes various missions whereby the sun or a laser source is used to drive a sailcraft. These range from Earth-orbiting satellites, station-keeping at a Lagrange point, and fly-by missions to inner and outer planets, asteroids, comets, and distant stars [6,7]. International and citizen space agencies have recently prepared early-stage demonstration missions such as IKAROS (JAXA), NanoSail D (NASA), Lightsail (Planetary Society), marking an age akin to the early days of air flight.

Juxtaposed with this interest is a proliferation of research on advanced diffractive materials. These include polarization diffraction gratings and other metamaterials having engineered optical properties [8-23]. Films of these materials may, in principle, be thin (a few micrometers thick) with a low areal density, and produced across large areas, thereby affording passive or active control of a sailcraft. We propose that diffractive materials may be used in place of reflective sails to afford non-mechanical navigation (e.g., without varying the sailcraft attitude), photon re-cycling, and higher efficiencies. For example, binary switching between the +1 and -1 order of an electro-optically controlled diffractive film may provide continuous navigational control authority. This report advances that vision by comparing the radiation pressure on reflective and diffractive surfaces, and applying the force models to a two-dimensional orbit-raising scenario. The paramount finding is that no penalty is found for the areal density and momentum transfer. A complete optimization analysis based on mission-specific criteria, not presented here, is left for further work. Polarization, dispersion, space-qualified materials, and other practical concerns are also left for future development.

Radiation pressure is described in Section 2 for flat films. Four special cases are examined for light governed by the law of reflection and diffraction, including both reflection and transmission cases in the Littrow configuration, and a grating at normal incidence. A single high-efficiency diffraction order is assumed. A two-body point mass model comprised of the sail and the Sun in a two-dimensional orbital plane is described in Section 3. The force model is applied in Section 3(A) to raise the sailcraft from an Earth orbit to that of Mars. Section 3(B) outlines advanced diffractive sail control schemes. Concluding remarks are offered in Section 4.

## 2. RADIATION PRESSURE ON A FLAT SAIL

A uniform plane wave propagating along the incident unit vector direction $\hat{k}_i$ with irradiance $I_i$ is assumed. The incident force on a flat film of surface area $A$ scales with the collected attitude-dependent incident beam power $P_i(\theta_i)$:

$$\vec{F}_i = P_i(\theta_i)\hat{k}_i / c = (I_i A / c)\hat{k}_i \cos\theta_i \tag{1}$$

where $c$ is the speed of light, $\theta_i$ is the angle subtending the incident wave vector and the outward surface normal of the front face $\hat{n}_f$ (see Fig. 1). For sailcraft moving much slower than the speed of light the Doppler shift of reflected and transmitted waves may be ignored, and thus the magnitudes of the wave vectors are equal: $k_i = k_r = k_t$. For a non-absorbing film of reflectance $R$, transmittance $T$, with $R+T = 1$, the net force may be expressed:

$$\vec{F}_{rp} = \vec{F}_i + \vec{F}_r + \vec{F}_t = (I_i A / c)\cos\theta_i \left( \hat{k}_i - R\hat{k}_r - T\hat{k}_t \right) \tag{2}$$

Corrections to Eq. (2) are required if a significant fraction of the light is diffusely reflected or transmitted, or if heat is emitted owing to absorption. Here we seek to investigate ideal cases, leaving higher order material-specific complications for later explorations.

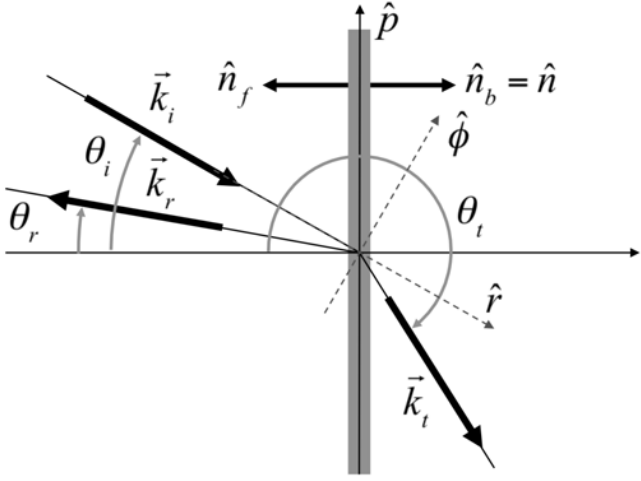

Fig. 1. Plane of incidence for rays at a flat planar film. Film coordinates $(n,p)$ and sunline coordinates $(r,\phi)$ are shown.

## A. Force components in the film coordinate system $(n,p)$

If the reflected and transmitted rays are constrained to the plane of incidence, the force may be projected along two unit vectors: normal and parallel to the film with respective unit vectors $\hat{n}$, $\hat{p}$. The ray directions may then be expressed (see Fig. 1):

$$\hat{k}_i = \cos\theta_i \hat{n} - \sin\theta_i \hat{p} \ , \ \ \hat{k}_{r,t} = -\cos\theta_{r,t}\hat{n} + \sin\theta_{r,t}\hat{p} \qquad (3)$$

where we assign $\hat{n} = \hat{n}_b = -\hat{n}_f$. The three radiation pressure forces may then be expressed:

$$\vec{F}_i = F_0 \cos\theta_i \hat{k}_i = F_0\left(\cos^2\theta_i\hat{n} - \cos\theta_i\sin\theta_i\hat{p}\right)$$
$$\vec{F}_r = -F_0 R\cos\theta_i \hat{k}_r = F_0 R\cos\theta_i\left(\cos\theta_r\hat{n} - \sin\theta_r\hat{p}\right)$$
$$\vec{F}_t = -F_0 T\cos\theta_i \hat{k}_t = F_0 T\cos\theta_i\left(\cos\theta_i\hat{n} - \sin\theta_i\hat{p}\right)$$

$$\qquad (4)$$

where the scaling force $F_0 = (I_i A / c)$ is defined for convenience. For example, at 1 AU the solar irradiance is 1.37 [kW/m²] and thus $F_0 = 4.57$ [$\mu$N] for a square meter sail. Below we shall consider either a perfecting reflecting or perfectly transmitting film (notated by $r$ or $t$). In such cases the normalized force may be expressed as an efficiency vector $\vec{\eta} = \vec{F}_{rp} / F_0$ :

$$\vec{\eta}^{r,t} = \cos\theta_i\left((\cos\theta_i + \cos\theta_{r,t})\hat{n} - (\sin\theta_i + \sin\theta_{r,t})\hat{p}\right) \qquad (5)$$

## B. Force components in the orbital coordinate system $(r,\phi)$

The sunline $\hat{k}_i$ is assumed to radiate from a point-like sun, parallel to the orbital radial unit vector $\hat{r}$ in a two dimensional plane: $\hat{r} = \hat{k}_i = \cos\theta_i\hat{n} - \sin\theta_i\hat{p}$ and $\hat{\phi} = \sin\theta_i\hat{n} + \cos\theta_i\hat{p}$. The angle of incidence $\theta_i$ describes the attitude of the film with respect to the sunline. From Fig. 1 we see $\hat{x} = \hat{k}_i = \cos\theta_i\hat{n} - \sin\theta_i\hat{p}$ and $\hat{y} = \sin\theta_i\hat{n} + \cos\theta_i\hat{p}$ and thus we express the components of Eq.s (5)

$$\eta_r^{r,t} = \vec{\eta}_{rp}^{r,t} \cdot \hat{r} = \cos\theta_i\left(1 + \cos(\theta_{r,t} - \theta_i)\right)$$
$$\eta_\phi^{r,t} = \vec{\eta}_{rp}^{r,t} \cdot \hat{\phi} = -\cos\theta_i\sin(\theta_{r,t} - \theta_i)$$

$$\qquad (6)$$

## C. Special Cases

Several cases of special interest may be described, including films that obey the Reflection Law, diffraction from a grating at the Littrow condition for both reflection and transmission, and, diffraction from a grating for the case of normal incidence.

i) Reflection Law (RL). $\theta_r = -\theta_i$ , $\left|\theta_r\right| \le \pi / 2$

$$\vec{\eta}^{RL} = 2\cos^3\theta_i\hat{n} = 2\cos^2\theta_i\left(\cos\theta_i\hat{r} - \sin\theta_i\hat{\phi}\right) \qquad (7)$$

ii) Littrow Reflection (LR) Grating. $\theta_r = \theta_i$ , $\left|\theta_r\right| \le \pi / 2$

$$\vec{\eta}^{LR} = 2\cos\theta_i\left(\cos\theta_i\hat{n} - \sin\theta_i\hat{p}\right) = 2\cos\theta_i\hat{r} \qquad (8)$$

iii) Littrow Transmission (LT) Grating. $\theta_t = \pi - \theta_i$ , $\left|\theta_t\right| \ge \pi / 2$

$$\vec{\eta}^{LT} = -2\cos\theta_i\sin\theta_i\hat{p}$$
$$= 2\cos\theta_i\sin\theta_i\left(\sin\theta_i\hat{r} - \cos\theta_i\hat{\phi}\right) \qquad (9)$$

iv) Grating at normal incidence : $\theta_i = 0$, $\hat{n} = \hat{x}$, $\hat{p} = \hat{y}$

$$\vec{\eta}^{NI} = \left((1 + \cos\theta_{r,t})\hat{r} - \sin\theta_{r,t}\hat{\phi}\right) \qquad (10)$$

The force is directed along the surface normal of the back face for a reflective film, but the force is parallel to the film for a transmission grating at the Littrow condition. A reflection grating at the Littrow condition is the vector sum of these two values, with a magnitude equal to $2F_0\cos\theta_i$.

The components of force in both the $(n,p)$ and $(r,\phi)$ basis are plotted for these three cases in Fig. 2(a,b) as a function of the angle of incidence. We see from Eq. (8) and from Fig. 2(b) that as the attitude of a Littrow reflection grating changes, the force remains directed only along the sunline. This, and the relatively larger force, may provide a dynamic advantage in some cases (for example, an oscillating attitude does not produce an undulating trajectory). In contrast, an orbit-raising maneuver may require a significant transverse lift force with $F_\phi \ge F_r$.

In those cases a reflective mirror or a transmissive grating at the Littrow condition may be desired. However, those two cases do not constitute the optimal lift force, owing the weakening factor of $\cos\theta_i$ in both Eq. (7) and (9). A diffractive film illuminated at normal incidence overcomes this disadvantage. In this case the magnitude of force for a normally incident grating is $2^{3/2}F_0\cos(\theta_{r,t} / 2)$, providing the maximum amount of force at any diffraction angle. What is more, the lift force efficiency, $\eta_\phi^{NI}$ may be as large at 100%.

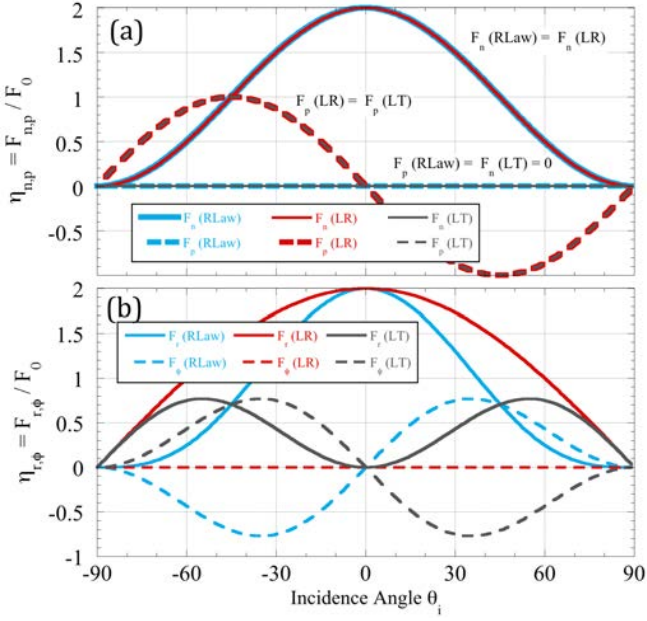

Fig. 2. Radiation pressure efficiency components projected along the body coordinates (n,p) and sunline coordinates (r, φ), plotted as a function of the angle of incidence, θ_i. (RLaw = Reflection Law, LR = Littrow Reflection, LT = Littrow Transmission)

Parametric force lines, shown in Fig. 3, provide a convenient way to represent the force components for the four cases examined above. A large magnitude of force and lift component are clearly afforded by the normally incident grating case.

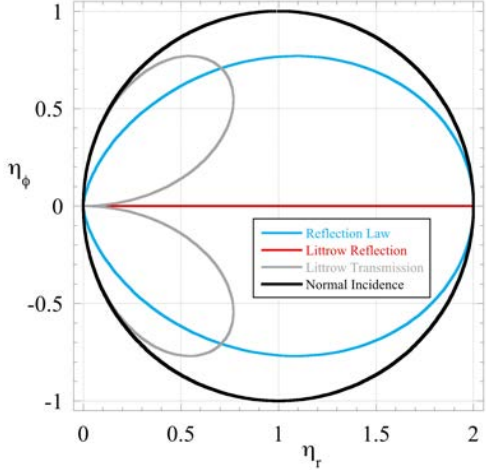

Fig. 3. Parametric force efficiency lines for reflective and diffractive films.

# 3. SOLAR PRESSURE ON AN ORBITING DIFFRACTIVE OR REFLECTIVE SAIL

The solar propulsion on a sailcraft of area $A$ may be estimated by assuming point-like masses for the sailcraft, $m$, and sun, $M$, and ignoring other gravitational bodies. Further simplicity is achieved by assuming a two dimensional sailcraft trajectory in the $xy$-plane ($r, \phi$-plane), with the sun at the origin. The net force from gravity and radiation pressure may be expressed [6]

$$\vec{F} = \vec{F}_G + \vec{F}_{RP} = -G\frac{mM}{r^2}\hat{r} + A(I_E/c)(R_E/r)^2\vec{\eta} = m\vec{a} \quad (11)$$

where $I_E = 1.37$ [kW/m²] is the so-called solar irradiance constant at $r = R_E = 1$ [AU], $G$ is the universal gravitational constant, and $\vec{\eta} = \vec{F}_{rp}/F_0$ is the radiation pressure efficiency vector. The acceleration may be expressed in circular coordinates:

$$\vec{a} = \frac{d^2\vec{r}}{dt^2} = \left(\ddot{r} - r\dot{\phi}^2\right)\hat{r} + \left(2\dot{r}\dot{\phi} + r\ddot{\phi}\right)\hat{\phi} \quad (12)$$

Expressing the area mass density $\sigma = m/A$, Eq. (11) may be written:

$$\vec{F} = -ma_M(R_E/r)^2\left((1 - \eta_r\sigma_{cr}/2\sigma)\hat{r} - (\eta_\phi\sigma_{cr}/2\sigma)\hat{\phi}\right) \quad (13)$$

where $\sigma_{cr} = 2R_E^2 I_E/GMc = 1.54\,g/m^2$ is a characteristic mass density and $a_M = -GMR_E^{-2} = 5.931$ [mm/s²] is the gravitational acceleration of the sun at the radial orbit of the Earth. The ratio $\sigma^* = \sigma_{cr}/\sigma$ is called the lightness number. Note that the sailcraft is neutrally buoyant with respect to solar gravity when $\sigma^* = 1$, assuming $\eta_r \gtrsim 2$ (a sun-facing mirror or retro-reflecting grating). The values of the radiation pressure terms in Eq. (13) tend toward infinity as the areal density of the sailcraft vanishes ($\sigma \to 0$). With current technology a modest value of $\sigma^* = 0.1$ is within reach for CubeSat compatible sailcraft for example.

## A. Synchronous Transfer Orbits with Constant Solar Attitude

To demonstrate that there is no obvious advantage of reflective over diffractive sailcraft, let us examine a simplified Earth to Mars rendezvous mission. The trajectory will be a smooth outward spiral, as shown in Fig. 3. For convenience, we assume the planets have circular co-planar orbits of respective radii $R_E = 1.0$ [AU] and $R_M = 1.5$ [AU]. The initial ($t = 0$) and final ($t = T$) conditions for the rendezvous are $r(t=0) = R_E$, $\vec{v}(t=0) = v_E\hat{\phi}$, $r(t=T) = R_M$, and $\vec{v}(t=T) = v_M\hat{\phi}$, where $v_E = \sqrt{GM/R_E}$ and $v_M = \sqrt{GM/R_M}$. The desired value of the radial component of velocity at the two time points is zero. Lacking a general analytic solution, this non-central potential type problem may be conveniently solved by numerical means (e.g., 4th order Runge-Kutta). As a further simplification, let us assume a fixed sailcraft attitude with respect to the sunline and a fixed diffraction or reflection angle throughout the orbit. The sail may be jettisoned or stowed once the desired orbit is reached. The numerical challenge then is to determine the diffraction or reflection angle that satisfies the boundary conditions (to within a small error) in the shortest time, $T$. In principle, an exact matching of the boundary conditions may not exist for this type of orbit. Nevertheless, they may be satisfied to within a given degree of error. In the case of a fixed solar attitude, we achieved quasi-synchronous transfers with errors for terminal radius, energy, and azimuthal velocity below 0.01% of the expected values for Mars. The gravitational sphere of influence of Mars is 170 radii, or 0.25% of the orbital radius, hence, an error < 0.01% is akin to a bull's eye. The error for the radial component of velocity, however, was typically several tenths of a percent. This suggests that a fixed attitude sailcraft cannot make a perfectly synchronous transfer unless either the initial condition is changed by the boost-phase rocket from Earth, or if the components of radiation pressure are actively controlled during the mission [24,25].

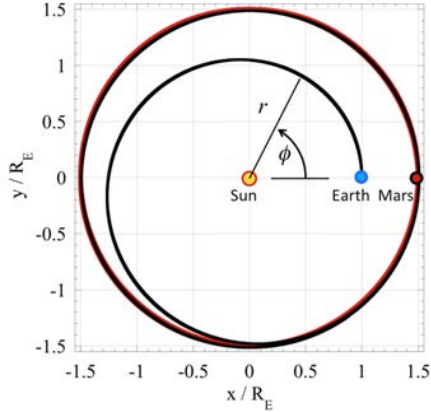

Fig. 4. Typical synchronous Earth-Mar transfer orbit for a fixed solar attitude (lightness number σ* = 0.10). Sail is jettisoned upon reaching the desired orbit.

### 1. Ideal Mirror (Case i)

At a lightness value of $\sigma^* = 0.1$ we find a numerical solution for a matched Earth to Mars orbit occurs when $\theta_i = 50°$, resulting in a transfer time of $T$=1.58 years. As the lightness increases value of $\theta_i$ must also change to satisfy the orbital boundary conditions, as show in Fig. X(a). Reducing the areal density of the sailcraft, however, does not significantly change to time to reach the matched orbit. As shown in Fig. X(b), the azimuthal acceleration term in Eq. (13), $\eta_\phi \sigma_{cr} / 2\sigma$, remains relatively constant as $\sigma^*$ is varied, while the radial value, $\eta_r \sigma_{cr} / 2\sigma$, decreases as the sailcraft become more buoyant against the sun. We note that a matched orbit cannot be achieved unless $\sigma^* > \sim 0.08$ and $\theta_i \leq 45°$. As verified by the cases below, the cut-off condition for $\sigma^*$ occurs when $\eta_r = \eta_\phi$ or equivalently. Therefore we see that the lift force must be greater than or equal to the radial force from radiation pressure. The relative invariance of the time of arrival offers no compelling need to reduce the areal density of the sailcraft below $10\sigma_{cr}$, assuming the mission flight path allows an incidence angle of roughly 50°. Of course, other mission factors not considered here may benefit from lower areal densities.

### 2. Littrow Reflection Grating (Case ii)

Lacking azimuthal acceleration, there are no solutions for a Littrow reflection grating.

### 3. Littrow Transmission Grating (Case iii)

For a transmissive Littrow grating of lightness $\sigma^* = 0.1$ a Mars orbital condition is reached in $T$=1.44 years when $\theta_i = 21.5°$. A second solution is also predicted: $T$=1.52 years when $\theta_i = 50.7°$. If the lightness value increases, as shown in Fig.x (a), we find that the shorter time branch remains relatively constant, as found in Case (i), whereas the longer time branch increases linearly with $\sigma^*$. The cut-off at time branch remains relatively constant, as found in Case (i), whereas the longer time branch increases linearly with $\sigma^*$. The cut-off at the longer time branch increases linearly with $\sigma^*$. The cut-off at $\sigma^* = 0.08$ and $\theta_i = 45°$ is seen again, with the lower short-time branch (upper long-time branch) corresponding to $\eta_r < \eta_\phi$ ($\eta_r > \eta_\phi$), as evident in Fig.X(b). As for the case of a reflective sail, a smaller radial component of radiation pressure force, compared to the azimuthal lift component is favorable for a short arrival time. What is more, from a materials fabrication point of view it may be desirable to require a small Littrow angle, which in principle is easier to achieve. This latter point also benefits from a larger lightness value, since $\theta_i$ becomes vanishingly small as $\sigma^*$ approaches unity. As shown in Fig. X the azimuthal values $\eta_\phi \sigma_{cr} / 2\sigma$ are found to be nearly invariant with lightness. For example, a value of $\theta_i = 9.4°$ is matched to $\sigma^* = 0.2$, providing a transit time of $T$=1.42 years.

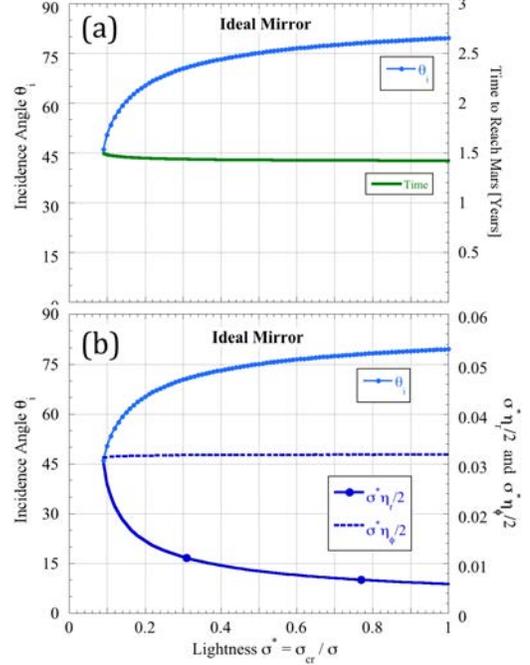

Fig. 5. Quasi-synchronous transfer orbit for a mirror-based sailcraft. Optimized attitude, $\theta_o$ as a function of lightness number $\sigma^*$. Also plotted are the corresponding transit time (a), and acceleration factors (b).

### 4. Sun-Facing Grating (Case iv)

A transmissive sailcraft propelled by a sun-facing ($\theta_i = 0$) diffractive sail provides a transit time of $T$=1.44 years if $\theta_t = 141°$ ($\theta_t' = 180° - \theta_t = 39°$) and $\sigma^* = 0.1$. A reflective branch was also found, but the transit time was greater than two years across a large range of lightness values, so it is not analyzed here. The transfer time for the transmissive branch is relatively invariant to lightness, as seen in cases above. The required value of $\theta_t'$ decreases with increasing lightness. For example, at $\sigma^* = 0.2$ a value of $\theta_t' = 18.6°$ is required, providing a journey of $T$=1.42 years. The cut-off angle corresponding to $\eta_r = \eta_\phi$ is $\theta_t = 90°$. From Fig.7(b) we again see that $\eta_r < \eta_\phi$ is required, and $\eta_\phi \sigma_{cr} / 2\sigma$ is relatively invariant to lightness.

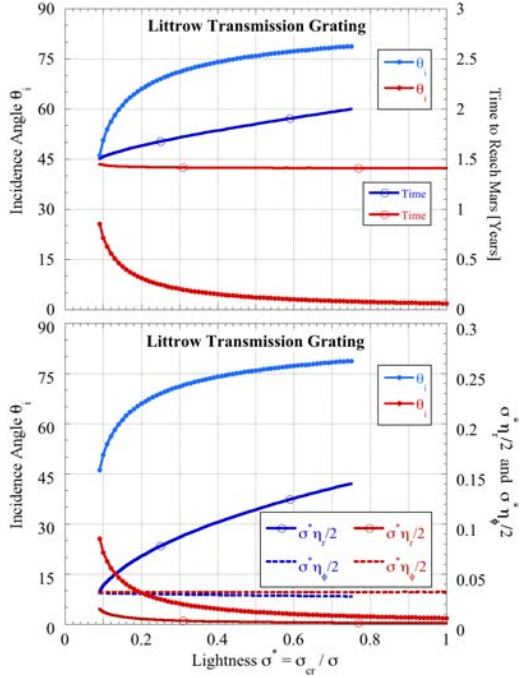

Fig. 6. Quasi-synchronous transfer orbit for a Littrow transmission grating sailcraft. Optimized attitude, $\theta_i$, as a function of lightness number $\sigma^*$. Also plotted are the corresponding transit time (a), and acceleration factors (b). Red-colored branches corresponding to the shortest transit times.

## 5. Comparisons

The ideal mirror requires 10% more time to reach a Martian orbit with a fixed attitude sailcraft. This difference may be expected to significantly vary if the attitude is actively controlled or if a diffractive metasail includes an electroptic control mechanism. For example, an optimally controlled reflective sail mission to Mars provided predicted a transfer in as short as 324 days [24,25]. At a lightness number $\sigma^* = 0.1$ we found the Littrow transmission case requires an attitude inclination of $\theta_i = 21.5°$, and thus, a total beam deviation of twice this value. In comparison the sun-facing grating must deviate the transmitted beam by $\theta'_t = 39°$. These angular deviation values are comparable, and thus neither configuration is most favorable from the point of view of engineering the diffractive film. On the other hand, smaller deviations angle may be easier to fabricate – especially if uniform broadband performance is preferred, as it lessens the deviation angle. Finally is should be noted that the Earth-Mars transfer times in this report are longer than the chemically fuel Hohmann transfer, which is roughly $T$=0.7 years. It remains an open question weather an actively controlled solar sail can achieve shorter times.

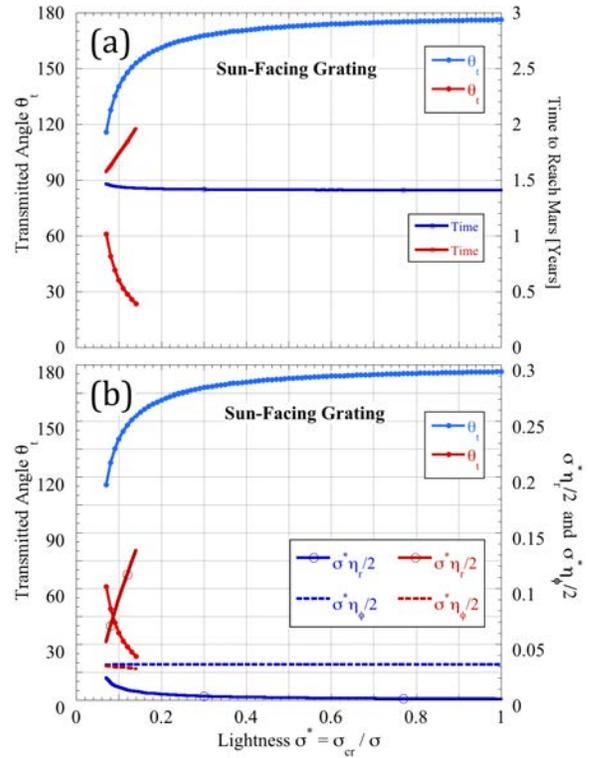

Fig. 7. Quasi-synchronous transfer orbit for a sun-facing diffractive sailcraft. Optimized attitude, $\theta_i$, as a function of lightness number $\sigma^*$. Also plotted are the corresponding transit time (a), and acceleration factors (b). Blue-colored branches corresponding to the shortest transit times.

## B. Controlled Binary Metamaterial Arrayed Grating

As an example of an electro-optically controlled diffractive sailcraft, let us chose a sun-facing sail with diffractive panels that are switchable between two equal but opposite diffraction orders: $\pm\theta_1$. Note that the radial force on the sailcraft does not change when the diffraction order is switched, whereas the lift component changes sign. For a large number of such panels arrayed across the sail, the average azimuthal force may be varied nearly continuously between two equal but opposite values. Combining Eq.s (10) and (13) we write

$$\vec{F} = -ma_M (R_E/r)^2 \left( (1 - \frac{\sigma^*}{2}(1 + \cos\theta_1))\hat{r} + g(t)\frac{\sigma^*}{2}\sin\theta_1\hat{\phi} \right)$$

where $g(t)$ is the control variable with $|g(t)| \leq 1$ and defined assuming $\theta_1 > 0$. For example, $g(t)$ has a negative value when most of the panels diffract into the $-\theta_1$ order. For a reflective grating, $0° \leq \theta_1 \leq 90°$, whereas $90° \leq \theta_1 \leq 180°$ for a transmissive grating (see Fig. 1). If we require $\eta_\phi > \eta_r$ during some part of the transfer, as was necessary for short transfer times in Section A, then a transmissive grating is required. The determination of a control signal that achieves a synchronous transfer is the shortest time is beyond the scope of this report, and will be explored in future studies.

## 4. CONCLUSIONS

Diffractive and reflective sails having a fixed attitude with respect to the sunline were examined. Nearly synchronous Earth-Mars transfers were achieved, with short transfer times when the transverse lift force exceeded the radial scattering force from solar radiation pressure. A transmissive sun-facing diffractive sail composed of an array of switchable diffractive elements is of particular interest, affording a variable lift component of force when the sign of the diffractive order is switched. Future work is needed to find an optimized control scheme for such a sailcraft.

**Funding Information.** National Science Foundation (NSF): ECCS-1309517.